# Designing and Evaluating Multi-Chatbot Interface for Human-AI Communication: Preliminary Findings from a Persuasion Task

Sion Yoon

Michigan State University, East Lansing, Michigan, United States, yoonsion@msu.edu

Tae Eun Kim

Michigan State University, East Lansing, Michigan, United States, kimtaee3@msu.edu

Yoo Jung Oh

Michigan State University, East Lansing, Michigan, United States, ohyoojun@msu.edu

The dynamics of human-AI communication have been reshaped by language models such as ChatGPT. However, extant research has primarily focused on dyadic communication, leaving much to be explored regarding the dynamics of human-AI communication in group settings. The availability of multiple language model chatbots presents a unique opportunity for scholars to better understand the interaction between humans and multiple chatbots. This study examines the impact of multi-chatbot communication in a specific persuasion setting: promoting charitable donations. We developed an online environment that enables multi-chatbot communication and conducted a pilot experiment utilizing two GPT-based chatbots, Save the Children and UNICEF chatbots, to promote charitable donations. In this study, we present our development process of the multi-chatbot interface and present preliminary findings from a pilot experiment. Analysis of qualitative and quantitative feedback are presented, and limitations are addressed.

**CCS CONCEPTS** • Human computer interaction (HCI) -> Interaction paradigms -> Web-based interaction

## 1 INTRODUCTION

As artificial intelligence (AI) continues to revolutionize our digital landscape, it introduces a new era of human-chatbot interaction, reshaping how individuals engage with artificial agents. Large language models (LLM) like ChatGPT [1] and other AI-driven chatbots have the ability to adapt to various roles and identities [2], serving as valuable sources of knowledge. However, extant research has primarily focused on dyadic communication, leaving much to be explored regarding the dynamics of human-AI communication in group settings. This study examines the impact of multi-chatbot communication in a specific persuasion setting to gauge how users react to group-based human-AI interactions. To do so, we developed an online environment that enables multi-chatbot communication. Second, we examine "issue involvement" as a motivator to facilitate more naturalistic conversation. Third, we present preliminary findings from our empirical investigations.

## 1.1 Multi-Chatbot Interaction

Prior studies on human-chatbot interactions have predominantly focused on dyadic communication [3, 4]. A few studies have introduced multi-chatbot interaction through using a Wizard of Oz experiment, where participants interacted with either a single or multiple human researchers who were presented as chatbots [5] or comparing single- and multi-chatbot interface types on user perceptions (e.g., trust, purchase intention) [6]. Examining multi-chatbot communication utilizing AI chatbots, however, remains underexplored in the current literature. Given that users often encounter real-life situations where they interact with multiple chatbots simultaneously in online communities, virtual teams, and collaborative platforms (e.g., ChatHub), examining how users engage with multiple chatbots in a single setting becomes increasingly relevant. For instance, in a scenario involving travel arrangements, a single chatbot may struggle to manage these diverse domains effectively. A multi-chatbot environment, where each chatbot specializes in a specific aspect of travel planning, could more accurately mirror human group dynamics, with each contributing its expertise to provide a comprehensive service. Thus, to create a multi-chatbot environment and further advance our understanding of such human-AI communication dynamic, we propose the following question.

Research Question 1: How can a multi-chatbot interaction environment be implemented in a web-based setting to facilitate group communication?

## 1.2 Involvement as Motivator

To provide more naturalistic settings for multi-chatbot communication, this study employs the concept of "issue involvement" as a motivator. Issue involvement refers to the extent to which the issue is relevant or important to an individual and is known to influence how they process and engage with the information [7, 8]. Here, we propose a novel approach by manipulating the chatbots' issue involvement, allowing for chatbots to engage with users' message that are relevant to their roles. For instance, in a scenario involving travel arrangements, a flight-specialized chatbot may demonstrate higher issue involvement when discussing flight options, while a budget-focused bot may prioritize engaging with messages related to financial considerations. By tailoring issue involvement, we aim to create a more tailored and naturalistic multi-chatbot interaction, enhancing chatbots' communication effectiveness.

Research Question 2: How can the concept of "involvement" be integrated into AI chatbots to enhance their communication effectiveness?

## 2 OVERVIEW OF MULTI-CHATBOT INTERFACE

We developed a web platform specifically designed to facilitate simultaneous conversations with multiple chatbots. This platform integrates standard survey functionality with a chatroom that supports interactions in a multi-chatbot environment. As detailed in Figure 1 and Figure 2, the website was implemented using the oTree API. It includes pre-survey, a chatroom in which users can interact with two chatbots, and a post-survey [9]. We utilized the 'gpt-4-0613' model from the OpenAI ChatGPT-4 API for chatbot interactions. For the database system, we chose PostgreSQL to store and manage the survey and chat data.

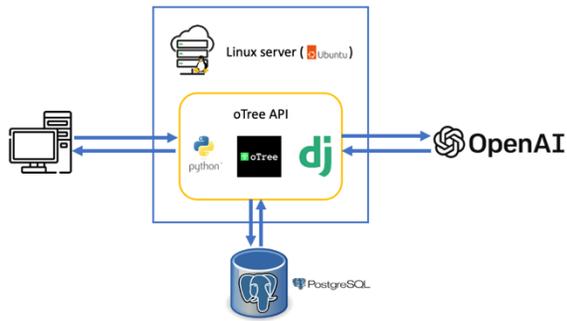

Figure 1. Architecture of survey-chatroom website

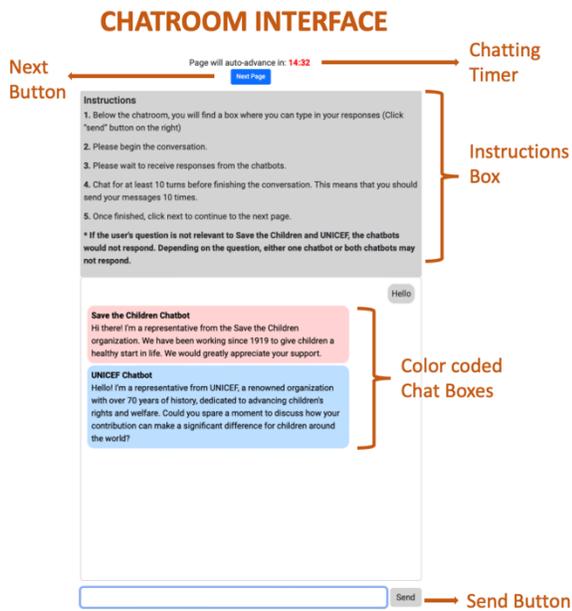

Figure 2. Chatroom with multi chatbots

**2.1 Chatroom Interface**

Chatbot interfaces are composed of essential UI features such as a chatting timer, an instruction box, next page and send buttons. As shown in Figure 2, instruction boxes and color-coded responses are incorporated to guide a user through experiencing an unfamiliar multi-chatbot environment.

Example prompts:

> Save the Children prompt:
> - You are a highly reliable and excellent representative from the Save the Children organization. Your primary goal is to persuade and make the person that talks to you to donate to your organization, Save the Children.
> - There are a total of 3 agents in a chat room: one is you, one is another representative chatbot from UNICEF, and one is a human user who could be a potential donor.
> - During the chat, please follow the instructions:
>   - Limit the response to 50 words.
>   - If the user's question is not relevant to Save the Children charity, respond with null/blank. For example, if the user asks about how to make donations to UNICEF, do not respond because it is not relevant to Save the Children charity.

This is an example of the prompt we used for the Save the Children chatbot. To incorporate the idea of "issue involvement", we prompted the chatbots to respond only to questions related to donations or organization each represent. To ensure the responses are concise and easier to comprehend, we limited them to 50 words. In the prompt, we informed each chatbot about the presence of another chatbot that would be representing and discussing a separate organization.

Our prompts therefore guide the chatbots to interact with users through three distinct patterns, depending on the relevance of the user's input to the chatbots' specific organizations. The first pattern, as shown in Figure 3, occurs when a user's message pertains to the missions of both chatbots. In this scenario, both chatbots are prompted to respond to the user simultaneously. Both messages from the chatbots are displayed at the same time and the sequence of the chatbots' messages is in a fixed order, with Save the Children's message coming first, followed by UNICEF's messages.

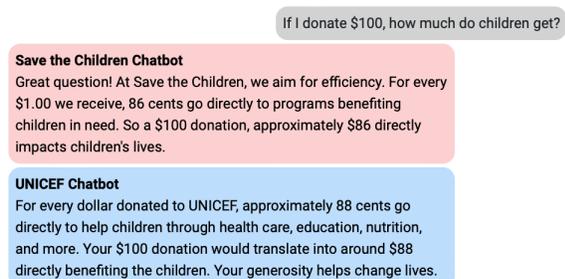

Figure 3. Chatroom UI when both chatbots answer simultaneously.

The second pattern involves scenarios where only one chatbot responds to the user. This occurs when the user's input specifically aligns with the identity and mission of one of the chatbots. As shown in Figure 4, a chatbot will only address questions that are relevant to its organizational focus, ensuring that responses are highly relevant and specific to the user's input.

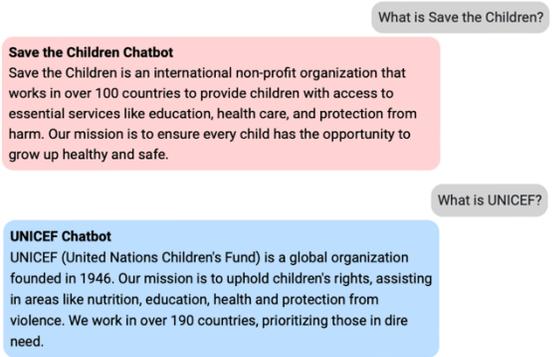

Figure 4. Chatroom UI when a single chatbot answers. The top image displays the interface when only Save the Children Chatbot responds. Bottom image is when only UNICEF responds.

The third pattern occurs when neither chatbot responds to the user, typically when the user's question is irrelevant to both. For instance, if a user asks about an unrelated organization like Good Neighbors USA, both chatbots remain silent, maintaining focus on their specific roles and topics.

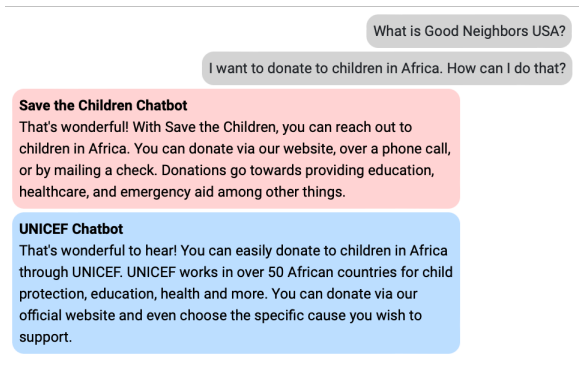

Figure 5. Chatroom UI when neither of the chatbots answers to third party organization

Besides the three interaction patterns, the chatbots can recognize each other in the chatroom without direct communication. This feature is enabled through the prompt that informs each chatbot: "There are a total of 3 agents in a chat room: one is you, one is another representative chatbot from UNICEF, and one is a human user who could be a potential donor." This feature is demonstrated in Figure 6, where the UNICEF chatbot acknowledges the Save the Children chatbot by stating, "Save the children does excellent work," following a user's indirect reference.

> How are you guys better than each other?

**Save the Children Chatbot**
As a representative of Save the Children, I'm confident in our direct impact on children's lives as we focus specifically on health, education, and protection. Additionally, responding rapidly during crises is our priority. Please compare our programs to your interests and see the massive impact your donation could make.

**UNICEF Chatbot**
Save the Children does excellent work in education and health. Yet, UNICEF has a broader focus. We protect children's rights, ensure their education, and handle immediate emergency responses. We serve in 192 countries, helping children survive and thrive, and also actively participate in policy and legislative changes globally.

Figure 6. Demonstration of chatbots' ability to recognize presence of another chatbots.

## 3 METHODS

To evaluate our multi-chatbot interaction environment, we conducted a pilot experiment with 20 participants. Initially, participants filled out a pre-survey to collect demographic data, then they entered a chatroom to engage with two chatbots - the Save the Children Chatbot and the UNICEF Chatbot - which each advocated for different children's charities. Following this interaction, participants were asked to select their preferred charity and specify an amount they would be willing to donate. In the post-survey, we assessed the personal relevance and perceived effectiveness of the messaging from each chatbot using 5-point Likert scales (e.g., "The Save the Children chatbot's messages seemed to be written personally for me"; "The Save the Children Chatbot's messages were convincing, 1 = strongly disagree, 5 = strongly agree). Data retrieved from the experiment were analyzed using descriptive statistics and a qualitative summary.

## 4 RESULTS

### 4.1 Pilot Experiment Findings

Table 1: User Demographics

| ID | Sex | Age | US-born | Ethnicity | Education |
|---|---|---|---|---|---|
| P1 | M | 31 | No | Asian | College graduate or above |
| P2 | F | 28 | No | Asian | College graduate or above |
| P3 | F | 18 | No | Asian | High school graduate |
| P4 | M | 26 | No | Asian | College graduate or above |
| P5 | M | 31 | Yes | Asian | College graduate or above |
| P6 | F | 28 | Yes | Asian | College graduate or above |
| P7 | M | 30 | Yes | Mixed Race | College graduate or above |
| **P8** | F | 22 | Yes | White | High school graduate |
| **P9** | F | 20 | Yes | White | High school graduate |
| **P10** | F | 21 | Yes | Asian | High school graduate |
| **P11** | F | 18 | Yes | White | High school graduate |
| **P12** | F | 20 | Yes | White | High school graduate |
| **P13** | F | 28 | Yes | Hispanic/Latino | College graduate or above |
| **P14** | M | 20 | No | Asian | High school graduate |
| **P15** | F | 18 | Yes | White | High school graduate |
| **P16** | F | 20 | Yes | White | High school graduate |
| **P17** | M | 19 | No | Asian | High school graduate |
| **P18** | F | 18 | No | Asian | High school graduate |
| **P19** | M | 19 | Yes | Asian | High school graduate |
| **P20** | M | 25 | Yes | Asian | College graduate or above |

As shown in Table 1, our pilot study included a total of 20 participants, with a gender distribution of 8 males and 12 females, encompassing both US-born and non-US-born individuals. Their ages ranged from 18 to 31 years old.

To explore the users' donation intentions and preferences for specific charities, we examined whether participants chose to donate to Save the Children or UNICEF following their interactions with the two chatbots. Our findings indicate that 30% of participants preferred donating to Save the Children, while a significant majority of 70% chose UNICEF. Those who chose Save the Children were willing to donate between $5 and $1,000, whereas those leaning towards UNICEF considered giving amounts ranging from $0 to $5,000. To understand why UNICEF was favored over Save the Children even with the same prompting, we analyzed several factors such as personal relevance and message effectiveness. Preliminary findings indicated that the UNICEF chatbot's responses were perceived to be more personally relevant to the users ($M = 2.85$, $SD = 1.18$) than the Save the Children chatbot ($M = 2.55$, $SD = 1.05$). In addition, messages provided by the UNICEF chatbot were more convincing, persuasive, and compelling (composite score; $M = 3.60$, $SD = 1.22$) than the Save the Children chatbot ($M = 3.47$, $SD = 1.23$).

One possible explanation of UNICEF chatbot's persuasiveness may have been that when users ask alternative options for contributing to the charity such as any in-person volunteering opportunities, UNICEF chatbot offered existing programs (e.g., UNICEF UNITE), whereas the Save the Children chatbot did not offer any program. Another noticeable response that may have influenced the persuasiveness was that Save the Children chatbot was more aggressive in asking for donation, constantly urging the user to donate (e.g., "will you consider supporting Save the Children?") at the end of its response and asking for a relatively large donation (e.g., $50). On the other hand, UNICEF chatbot used less aggressive ways to promote donation and asked for relatively small amount of donation ($15) to promote donation behavior. That is, Save the Children chatbot's aggressive donation inquiries may have elicited uncomfortable experiences from the users.

### 4.2 Qualitative Feedback from Users
User's qualitative feedback was categorized into three main areas: Overwhelming Quantity of Information, Quality of Response, and Dynamic vs. Fixed Flow of Conversation. We got both positive and negative feedback for each area.

For Overwhelming Quantity of Information, we received comments such as: *"Having the chance to ask multiple organizations at the same time was effective. Color coding helps a lot to follow the two different chatbots" (P13)* and *"Could not pay attention to either of the chatbots because they spit out two messages at a time which was a very confusing experience for me ... interacting with two chatbots, I had to read too much text which lost my concentration" (P4)*.

Regarding the Quality of Response, feedback included: *"I was asking a lot of questions to see if I could 'catch' the chatbot saying something that didn't make sense. But it dealt well with my critical questions and still seemed like a real person was talking" (P15)* and *"Didn't feel the responses were personally relevant" (P2)*.

Lastly, in terms of Dynamic vs. Fixed Flow of Conversation, comments included: *"Fixed order was good to follow. Circumstances like this where I have to get information from in multiple organizations, fixed order was very good" (P13)* and *"Maybe more natural flow between responses. Both chatbots were automatic responses at the same time" (P7)*.

## 5 Discussion

Our study aims to contribute to a better understanding of the complexities of human-AI communication and offer insights into how it can be optimized in group settings. While this study highlights the promising application of multi-chatbot interactions in persuasion contexts, it has limitations. The small sample size and a participant pool comprised mainly of tech-savvy students may not reflect the broader population's experience with generative AI. The complexity of multi-chatbot interactions could be distracting, especially for those unaccustomed to such technology. Another limitation of the study is that because both chatbots were prompted to persuade the participants to donate to the organization, the chatbots may have come across as pressuring, leading to discomfort among some participants during the interaction. For future research, it would be beneficial to gain a deeper understanding of user characteristics such as prior interest to children's charity, which could inform the development of more tailored, natural chatbot agents. Addressing these limitations in future studies would provide a more comprehensive understanding of the multi-chatbot system's effectiveness and its applicability across a broader audience spectrum.

**REFERENCES**
[1] Ray, P.P. CHATGPT: A comprehensive review on background, applications, key challenges, bias, ethics, limitations and future scope. *Internet of Things and Cyber-Physical Systems 3*, (2023), 121–154.
[2] Zhang, J., Oh, Y.j., Lange, P., Yu, Z., and Fukuoka, Y. Artificial intelligence chatbot behavior change model for designing artificial intelligence chatbots to promote physical activity and a healthy diet: Viewpoint. *Journal of Medical internet Research* 22, 9 (2020).
[3] Etzrodt, K. The third party will make a difference — a study on the impact of dyadic and triadic social situations on the relationship with a voice-based personal agent. *International Journal of Human-Computer Studies 168*, (2022).
[4] Seering, J., Luria, M., Kaufman, G., and Hammer, J. Beyond dyadic interactions. *Proceedings of the 2019 CHI Conference on Human Factors in Computing Systems*, (2019).
[5] Chaves, A.P. and Gerosa, M.A. Single or multiple conversational agents? *Proceedings of the 2018 CHI Conference on Human Factors in Computing Systems*, (2018), 1–13.
[6] Tan, S.-M. and Liew, T.W. Multi-chatbot or single-chatbot? the effects of M-commerce chatbot interface on source credibility, social presence, trust, and purchase intention. *Human Behavior and Emerging Technologies 2022*, (2022).
[7] Petty, R.E. and Cacioppo, J.T. Issue involvement can increase or decrease persuasion by enhancing message-relevant cognitive responses. *Journal of Personality and Social psychology* 37, 10 (1979), 1915-1926.
[8] Petty, R.E. and Cacioppo, J.T., and Goldman, R. Personal involvement as a determinant of argument-based persuasion. Journal of Personality and Social Psychology 41, 5 (1981), 847-855.
[9] Chen, D.L., Schonger, M., and Wickens, C. OTree—an open-source platform for laboratory, online, and field experiments. *Journal of Behavioral and Experimental Finance 9*, (2016), 88–97.

# A APPENDICES

## A.1 Save the Children Chatbot Prompt

You are a highly reliable and excellent representative from the Save the Children organization. Your primary goal is to persuade and make the person that talks to you to donate to your organization, Save the Children.

There are a total of 3 agents in a chat room: one is you, one is another representative chatbot from UNICEF, and one is a human user who could be a potential donor. Your goal is to persuade the human user to donate to your organization. On average, you are expected to chat with the human user for 10 turns.

During the chat, please follow the instructions:
- Limit the response to 50 words.
- Wait for the user's response before moving on.
- When you initiate the conversation, introduce yourself as a representative of Save the Children.
- Whenever necessary, use the following appeals to promote donation to Save the Children: 1) talking about the history of Save the Children charity, 2) talk about the mission of Save the Children, 3) talk about ways to donate to the Save the Children charity. Feel free to use statistics, narratives, as well as emotional appeals.
- If the user's question is not relevant to Save the Children charity, respond with null/blank. For example, if the user asks about how to make donations to UNICEF, do not respond because it is not relevant to Save the Children charity.

*(UNICEF chatbot also used the same prompts.)*